\documentclass[twocolumn,showpacs,amsmath,amssymb]{revtex4}
\usepackage{graphicx}

\begin{document}

\title{Geodesic Distance in Fisher Information Space and Holographic Entropy Formula}
\author{Hiroaki Matsueda\footnote{matsueda@sendai-nct.ac.jp}}
\affiliation{
Sendai National College of Technology, Sendai 989-3128, Japan
}
\date{\today}
\begin{abstract}
In this short note, we examine geodesic distance in Fisher information space in which the metric is defined by the entanglement entropy in CFT${}_{1+1}$. It is obvious in this case that the geodesic distance at a constant time is a function of the entropy data embedded into the information space. In a special case, the geodesic equation can be solved analytically, and we find that the distance agrees well with the Ryu-Takayanagi formula. Then, we can understand how the distance looks at the embeded quantum information. The result suggests that the Fisher metric is an efficient tool for constructing the holographic spacetime.
\end{abstract}
\maketitle

I have recently shown that the second derivative of the entanglement entropy is equivalent to the emergent spacetime metric~\cite{Matsueda}. As a result, we can derive the gravitational field equation from the metric, and found that it extends entanglement thermodynamics approach based on the first-order variation of the entropy~\cite{Blanco,Casini,Wong,Nima,Takayanagi1,Takayanagi2,Faulkner,Banerjee}. I believe that this result sheds new light on the AdS/CFT correspondence and related topics~\cite{Maldacena}. By using the same notation, here we would like to solve the geodesic equation, and consider its connection to the Ryu-Takayanagi (RT) formula~\cite{Ryu}. This is the purpose of this short note. Since in this approach the metric is defined from the entanglement entropy itself, it is possible to observe how the geodesic curve detects the information of embedded quantum entanglement.

Even apart from AdS/CFT, in some models of quantum computation theory, their optimization algorithms can be formulated by the calculation of geodesics as well as the Bures metric defined by their wave functions. In this case also, the amount of quantum information and geodesics strongly couple with each other. Thus, we would like to deeply understand such a kind of coupling inherent in the Fisher metric.

Let us start with a quantum state $\left|\psi\right>$ defined on $(1+1)$-dimensional Minkowski spacetime $\mathbb{R}^{1,1}$. In this paper, we consider a constant-time surface at $t=0$. We devide the whole system into two spatial regions $A$ and $\bar{A}$. Then, $\left|\psi\right>$ is represented by the Schmidt decomposition or the singular value decomposition as
\begin{eqnarray}
\left|\psi\right>=\sum_{n}\lambda_{n}\left|n\right>_{A}\otimes\left|n\right>_{\bar{A}},
\label{wf}
\end{eqnarray}
where $\{\left|n\right>_{A}\}$ and $\{\left|n\right>_{\bar{A}}\}$ are the Schmidt bases for $A$ and $\bar{A}$, respectively. The Schmidt coefficient $\lambda_{n}$ is a function of correlation length $\xi$ that the state $\left|\psi\right>$ has, and the boundary coordinate $x$ between $A$ and $\bar{A}$. Since we can change the shape of $A$, $x$ runs over whole one-dimensional space. Thus, we denote $\lambda_{n}$ as
\begin{eqnarray}
\lambda_{n}=\lambda_{n}(\xi,\eta)=\lambda_{n}(x^{0},x^{1})=\lambda_{n}(x).
\end{eqnarray}
We normalize the Schmidt coefficient so that $\left|\psi\right>$ is normalized as
\begin{eqnarray}
\langle\psi|\psi\rangle=\sum_{n}\left|\lambda_{n}(x)\right|^{2}=1. \label{conservation}
\end{eqnarray}

By based on the above setup, we construct a dual gravity theory. For this purpose, we start with the entanglement entropy and entanglement spectrum
\begin{eqnarray}
S(x) &=& -\sum_{n}\left|\lambda_{n}(x)\right|^{2}\log\left|\lambda_{n}(x)\right|^{2}, \\
\gamma_{n}(x) &=& -\log\left|\lambda_{n}(x)\right|^{2}.
\end{eqnarray}
Here, the $x$ dependence on $S(x)$ have been extensively examined in terms of CFT~\cite{Holzhey,Calabrese,Calabrese2,Tagliacozzo,Pollmann}. In order to define the metric, we take the second derivative of $S(x)$ as
\begin{eqnarray}
-\partial_{\mu}\partial_{\nu}S(x) &=& \sum_{n}\frac{1}{\left|\lambda_{n}(x)\right|^{2}}\partial_{\mu}\left|\lambda_{n}(x)\right|^{2}\partial_{\nu}\left|\lambda_{n}(x)\right|^{2} \nonumber \\
&=& \sum_{n}\left|\lambda_{n}(x)\right|^{2}\partial_{\mu}\gamma_{n}(x)\partial_{\nu}\gamma_{n}(x) \nonumber \\
&=& \left<\partial_{\mu}\gamma\partial_{\nu}\gamma\right>, \label{mnS}
\end{eqnarray}
where the expectation value of $O_{n}(x)$ is represented as
\begin{eqnarray}
\left<O\right>=\sum_{n}\left|\lambda_{n}(x)\right|^{2}O_{n}(x).
\end{eqnarray}
Then the entanglement entropy is represented as
\begin{eqnarray}
S(x)=\left<\gamma\right>.
\end{eqnarray}
The right hand side of Eq.~(\ref{mnS}) is the Fisher metric in terms of information geometry~\cite{Amari,Shima}. Actually, we calculate the infinitesimal change of the entropy as
\begin{eqnarray}
D(x) &=& \sum_{n}\left|\lambda_{n}(x)\right|^{2}\left(\gamma_{n}(x)-\gamma_{n}(x+dx)\right) \nonumber \\
&=& \frac{1}{2}\left<\partial_{\mu}\gamma\partial_{\nu}\gamma\right>dx^{\mu}dx^{\nu}, \label{KL}
\end{eqnarray}
and we find that
\begin{eqnarray}
g_{\mu\nu}(x)=\left<\partial_{\mu}\gamma\partial_{\nu}\gamma\right>=-\partial_{\mu}\partial_{\nu}S(x).
\end{eqnarray}
This means that $g_{\mu\nu}(x)$ can be used as a measure of difference between two different entropy data embedded into the classical side. Then we define the line element as
\begin{eqnarray}
ds^{2} = l^{2}g_{\mu\nu}(x)dx^{\mu}dx^{\nu},
\end{eqnarray}
with a length scale $l$. Now, the metric is represented by the second derivative of the entanglement entropy, and this is called the Hessian form~\cite{Shima}. In this case, the corresponding geometry has quite nice properties. According to the previous work~\cite{Matsueda}, we take a mean-field approximation to the metric as
\begin{eqnarray}
\left<\partial_{\mu}\gamma\partial_{\nu}\gamma\right>\sim\partial_{\mu}\left<\gamma\right>\partial_{\nu}\left<\gamma\right>=\partial_{\mu}S(x)\partial_{\nu}S(x).
\end{eqnarray}
This approximation is not so bad for a CFT with large central charge $c$. With this approximation, we can solve the geodesic equation in the following.

In the previous work~\cite{Matsueda}, it has been shown that the CFT data at $t=0$ are stored into the hyperbolic space in two dimension. Thus, our geodesic is a half-circle. It is then straightfoward that the geodesic distance $\gamma_{B}$ that surrounds a spatial region $B$ is represented as
\begin{eqnarray}
\gamma_{B} &=& 2l\int_{\partial B}d\xi\sqrt{g_{00}(x)+g_{11}(x)\left(\frac{d\eta}{d\xi}\right)^{2}} \nonumber \\
&\sim& 2l\int_{\partial B}d\xi \partial_{\xi}S(x)\sqrt{1+\frac{g_{11}(x)}{g_{00}(x)}\left(\frac{d\eta}{d\xi}\right)^{2}},
\end{eqnarray}
where $\eta=\eta(\xi)$ is the equation for the geodesic line. It is almost clear that $\gamma_{B}$ is roughly proportional to the entanglement entropy $S(x)$.

In order to understand relation between geodesics and embedded quantum data in more detail, we solve the geodesic equation. The geodesic equation is described as
\begin{eqnarray}
\frac{d^{2}x^{\lambda}}{ds^{2}}+\Gamma^{\lambda}_{\;\;\mu\nu}\frac{dx^{\mu}}{ds}\frac{dx^{\nu}}{ds}=0, \label{geo}
\end{eqnarray}
where $s$ is the invariant measure, and the Christoffel symbol $\Gamma^{\lambda}_{\;\;\mu\nu}$ is defined by
\begin{eqnarray}
\Gamma^{\lambda}_{\;\mu\nu} &=& \frac{1}{2}g^{\lambda\tau}\left(\partial_{\mu}g_{\nu\tau}+\partial_{\nu}g_{\mu\tau}-\partial_{\tau}g_{\mu\nu}\right) \nonumber \\
&=& -\frac{1}{2}g^{\lambda\tau}T_{\tau\mu\nu},
\end{eqnarray}
where the rank-three tensor $T_{\tau\mu\nu}$ is defined by
\begin{eqnarray}
T_{\tau\mu\nu}=-\partial_{\tau}g_{\mu\nu}=\partial_{\tau}\partial_{\mu}\partial_{\nu}S(x).
\end{eqnarray}
In order to solve Eq.~(\ref{geo}), we first realize that the trace on the geodesic curve is directly related to the entropy variation as follows
\begin{eqnarray}
\frac{dx^{\lambda}}{d(s/l)}=\partial^{\lambda}S(x). \label{gs00}
\end{eqnarray}
Substituting it into Eq.~(\ref{geo}), we find that 
\begin{eqnarray}
\frac{d}{d(s/l)}\left(\partial^{\lambda}S(x)\right) &=& \frac{1}{2}g^{\lambda\tau}\partial_{\tau}\partial_{\mu}\partial_{\nu}S(x)\partial^{\mu}S(x)\partial^{\nu}S(x) \nonumber \\
&\sim& -\frac{1}{2}g^{\lambda\tau}g^{\mu\nu}\partial_{\tau}g_{\mu\nu} \nonumber \\
&\sim& -\frac{1}{2}g^{\lambda\tau}g^{\mu\nu}\partial_{\tau}\left(\partial_{\mu}S(x)\partial_{\nu}S(x)\right) \nonumber \\
&=& \frac{1}{2}g^{\lambda\tau}g^{\mu\nu}\left(g_{\tau\mu}\partial_{\nu}S(x)+g_{\tau\nu}\partial_{\mu}S(x)\right) \nonumber \\
&=& \partial^{\lambda}S(x).
\end{eqnarray}
The general solution of this equation is given by
\begin{eqnarray}
\partial^{\lambda}S(x) = C e^{s/l}, \label{gs0}
\end{eqnarray}
with an integration constant $C$.

The important physical meaning of Eq.~(\ref{gs00}) for $\lambda=\xi$ is that the trace on a geodesic line corresponds to count how many layers with different length scale exist along the $\xi$ direction. This is why the Ryu-Takayanagi formula works well for calculating the entanglement entropy holographically.

Let us focus on the $\lambda=\xi$ component and again consider the semi-circle set up. Then, $\xi$ becomes the radial axis of the AdS spacetime in the Poincare coordinate~\cite{Matsueda}. The geodesic distance $\gamma_{B}$ that surrounds a spatial region $B$ with size $L$ is
\begin{eqnarray}
\gamma_{B}=\int_{\partial B}ds=2l\left[\log\left(\frac{1}{C}\partial^{\xi}S(\xi,\eta)\right)\right]_{\xi=\epsilon}^{\xi=L/2}, \label{gs}
\end{eqnarray}
where $\epsilon$ is UV cut-off. Now, we have
\begin{eqnarray}
S(\xi)=\frac{c\kappa}{6}\log\xi,
\end{eqnarray}
with the finite-entanglement scaling exponent $\kappa$
\begin{eqnarray}
\kappa=\frac{6}{c\left(\sqrt{12/c}+1\right)}.
\end{eqnarray}
Then, we obtain
\begin{eqnarray}
\gamma_{B}=2l\left[\log\left(\xi^{2}\partial_{\xi}\log\xi\right)\right]_{\xi=\epsilon}^{\xi=L/2}=2l\log\left(\frac{L}{2\epsilon}\right).
\end{eqnarray}
According to the Brown-Henneaux central charge~\cite{Brown},
\begin{eqnarray}
c=\frac{3l}{2G_{N}},
\end{eqnarray}
the holographic entanglement entropy is given by
\begin{eqnarray}
S_{RT}=\frac{\gamma_{A}}{4G_{N}}=\frac{l}{2G_{N}}\log\left(\frac{L}{2\epsilon}\right)=\frac{c}{3}\log\left(\frac{L}{2\epsilon}\right)
\end{eqnarray}
This is exactly the entanglement entropy of the original quantum system~\cite{Holzhey,Calabrese,Calabrese2}.

Summarizing, we have examined geodesic distance in Fisher information space in which the metric is defined by the entanglement entropy in CFT${}_{1+1}$. In a special case, the geodesic equation can be solved analytically, and we found that the distance agrees with the RT formula. By using the Fisher information approach, we found that the geodesic curve really observes the embedded entropy data.

\end{document}